\documentclass{elsart}

\usepackage{epsfig}

\begin{document}

\begin{frontmatter}

\title{Vacancy diffusion in the Cu(001) surface II: Random walk theory}

\author[IL] {E. Somfai\thanksref{ellaks.present.address}},
\author[KOL]{R. van Gastel},
\author[KOL]{S. B. van Albada},
\author[IL] {W. van Saarloos} and
\author[KOL]{J. W. M. Frenken}

\address[IL]{Universiteit Leiden, Instituut-Lorentz, PO Box 9506,
2300 RA Leiden, The Netherlands}
\address[KOL]{Universiteit Leiden, Kamerlingh Onnes Laboratory,
PO Box 9504, 2300 RA Leiden, The Netherlands}

\thanks[ellaks.present.address]{Corresponding author. Present address:
Department of Physics, University of Warwick, Coventry, CV4~7AL, U.K.}

\begin{abstract}
We develop a version of the vacancy mediated tracer diffusion model,
which follows the properties of the physical system of In atoms
diffusing within the top layer of Cu(001) terraces. This model differs
from the classical tracer diffusion problem in that (i) the lattice is
finite, (ii) the boundary is a trap for the vacancy, and (iii) the
diffusion rate of the vacancy is different, in our case strongly
enhanced, in the neighborhood of the tracer atom.  A simple continuum
solution is formulated for this problem, which together with the
numerical solution of the discrete model compares well with our
experimental results.
\end{abstract}

\end{frontmatter}

%%%%%%%%%%%%%%%%%%%%%%%%%%%%%%%%%%%%%%%%%%%%%%%%%%%%%%%%%%%%%%%%%%%%%%%%

\section{Introduction}
\label{sec:intro}

% * a bit about different diffusing atoms
% * tracer diffusion
% * summary of Raoul's paper:
%   - impurity In introduced at substitutional 1st layer sites
%   - motion followed by stm; mechanism: vmd
%   - results on jvd, waiting time, jump freq vs T

Diffusion is one of the most commonly observed stochastic processes.
Random walks, the paths of diffusing particles, are among the first
typical applications in probability theory textbooks. However, the
diffusing particles, or their paths, are often invisible, and their
effect can be observed only indirectly. This is the case for example in
vacancy mediated tracer diffusion, when on a lattice of atoms a
diffusing vacancy displaces atoms along its path. The process is
observed by following a labeled (or {\em tracer\/}) atom, whose steps
are slaved to the motion of the vacancy.

In this paper we treat theoretically the physical system of diffusing In
atoms in a Cu(001) surface layer, described in detail in paper I
\cite{gastelSS}. In section \ref{sec:tracer} we review the
literature about tracer diffusion. We describe our discrete model in
section \ref{sec:discrete}, the classical tracer diffusion problem
adapted to better suit our experimental system. Finally section
\ref{sec:continuum} gives a simple continuum formulation whose solution
allows one to fit the experiments without adjustable parameters.

We start with summarizing the experimental observations
\cite{gastelSS,gastel01,gastel00}. On a Cu(001) surface with up to
several hundred atomic spacing wide terraces, a few surface Cu atoms
were substituted with In atoms. Then the area was periodically imaged
with a scanning tunneling microscope (STM), and the position of the In
atoms was followed in time.

To our initial surprise, the In atoms do not move in a typical diffusive
way, but instead are stationary for some time, and then suddenly
(unresolvable by STM) they jump to a nearby lattice site, often up to 5
or more nearest neighbor spacings away from their original position.
These {\em long jumps\/} tend to happen at the same instant for
different In atoms in the imaged area. To explain this, we found that
the only mechanism permissible on physical grounds \cite{gastelSS} is
vacancy mediated tracer diffusion.

Based on Embedded Atom Model calculations \cite{eam84,finnis84} we
expect \cite{gastelSS} that surface vacancies at room temperature have a
low concentration, of the order $10^{-9}$ on Cu(001). They
both are created and recombine at steps, and stay in the top layer of a
terrace, as it is energetically unfavorable to dive deeper. From the
typical jump rate of $10^6$ Hz and terrace widths of a few hundred
atomic spacings, their lifetime is estimated to be at most of the order
of milliseconds. (Vacancies reaching the middle of the terrace have the
longest expected lifetime.) As the long jumps of In atoms are the effect
of a single vacancy, this short lifetime explains why the dynamics of
the jumps can not be resolved with STM (with imaging rate up to 10 Hz),
while the long waiting time between jumps, of the order of 10 s, enables
one to distinguish independent vacancies.

The measured waiting time between the jump events has an exponential
distribution, supporting the explanation of the jumps as independent
events. The distribution of the jump vectors, which has been measured,
will be compared to the numerical model in this paper.

%%%%%%%%%%%%%%%%%%%%%%%%%%%%%%%%%%%%%%%%%%%%%%%%%%%%%%%%%%%%%%%%%%%%%%%%

\section{Tracer diffusion}
\label{sec:tracer}

% * state our problem: boundaries etc
% * literature:
%   - BH first exact lattice soln for unbiased, per/inf bc
%     no absorption; separate problem etc
%   - Toro: solve differently
%   - Newman: continuum soln
%   - Benichou, Oshanin: introduce bias (still no absorption)
% * need for our model

The problem of vacancy mediated tracer diffusion was considered for a
long time
\cite{beijeren85,brummelhuis88,toroczkai97,newman99,benichou01}. It has
been solved first for the simplest case \cite{brummelhuis88}, when the
diffusion of the vacancy is unbiased (all diffusion barriers are equal;
the tracer atom behaves in the same way as the other atoms), the lattice
is two-dimensional and periodic or infinite.  There is a single vacancy
present, it takes a nearest neighbor step in a random direction at
regular time intervals, and has an infinite lifetime, as there are no
traps.  The solution is constructed by separating the motion of the
tracer and the vacancy. The correlation of the steps of the tracer atom
is calculated from the probability that the vacancy returns to the
tracer from a direction which is equal, perpendicular or opposite to its
previous departure. The distribution of the tracer atom spreads with
time, and on an infinite lattice for large times it approaches the
following functional form:
\begin{equation}
P_t(\mathbf{r}) = \frac{2(\pi-1)}{\log t} \, \mathrm{K}_0 \left(
\frac{r}{[\log t / (4\pi(\pi-1))]^{1/2}} \right) \,,
\end{equation}
where $\mathrm{K}_0$ is the modified Bessel function\cite{as}, time $t$
measures the number of steps of the vacancy, and at $t=0$ the vacancy is
near the tracer atom.  The non-Gaussian shape of the spatial
distribution function is typical for vacancy mediated diffusion.
The average displacement of the tracer particle
diverges with time. This is a direct consequence of the fact that two is
the marginal dimension for the return probabilities in the random walk
problem. For higher dimensional problems, the probability that the vacancy
returns to the tracer particle is less than unity and the average displacement
of the tracer particle remains finite. However, for dimensions equal to or
smaller than two, the vacancy always returns to the tracer particle and as
a consequence its displacement diverges.

The same problem has been solved in an alternative way for all
dimensions \cite{toroczkai97}. From this solution one can calculate the
number of tracer-vacancy exchanges up to time $t$: in two dimensions its
distribution is geometric, with mean $(\log t)/\pi$. The continuum
version of this problem has been considered as well in the form of an
infinite-order perturbation theory \cite{newman99}; the solution matches
the asymptotic form of the lattice model.

In a very recent study the lattice calculations were generalized to
biased diffusion \cite{benichou01}. The difference between the tracer
atom and the substrate atoms was taken into account by having different
vacancy-tracer and vacancy-substrate exchange probabilities, while the
rate of vacancy steps was kept constant.  Repulsive interaction reduces,
moderately attractive interaction increases the spreading of the tracer
distribution.

Although these exact solutions are closely related to our In/Cu(001)
system, the differences, e.g. in boundary conditions and vacancy
lifetime make a direct comparison with experiments impossible. For this
purpose we develop a model of tracer diffusion which includes the
essential properties of the experimental system.

%%%%%%%%%%%%%%%%%%%%%%%%%%%%%%%%%%%%%%%%%%%%%%%%%%%%%%%%%%%%%%%%%%%%%%%%

\section{Discrete model}
\label{sec:discrete}

% * specify full model
% * separate problem: {ret prob + recomb} + tracer diff
% * results for ret.prob.
% * tracer atom diff, modified model with special lattice
% * results, and P(0,0)
%   T-dep of jump length

In this section we describe a discrete model for In/Cu(001) diffusion,
solve it numerically, and present the results.

Our model is defined on a two-dimensional simple square lattice of size
$l \times l$, centered around the origin. This corresponds to the top
layer of a terrace of the Cu surface, with borders representing steps.
All sites but two are occupied by substrate atoms. At zero time the two
remaining sites are the impurity (or tracer) indium atom, which we
release at the origin, and a vacancy at position $(1,0)$. This
corresponds to the situation immediately after the indium atom has
changed places with the vacancy, e.g. for the first time.

\begin{figure}	%%% Fig barriers
\centering\epsfig{file=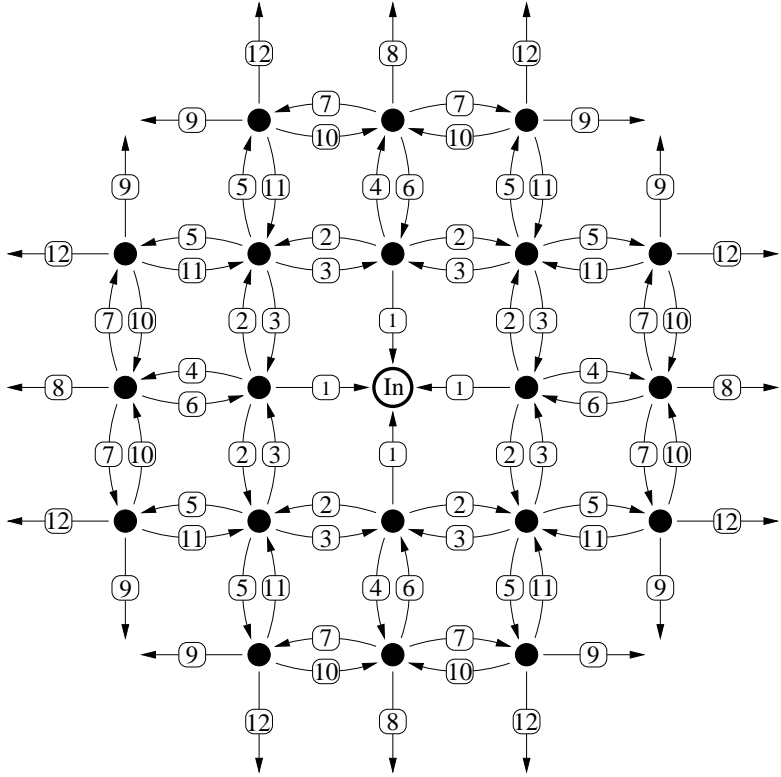}
\caption{Special vacancy diffusion barriers near an In atom, calculated
with the EAM method. The arrows denote the motion of the vacancy. The
following values for the barriers were used in the calculation:
$E_1   = 243$ meV,
$E_2   = 671$ meV,
$E_3   = 503$ meV,
$E_4   = 529$ meV,
$E_5   = 589$ meV,
$E_6   = 382$ meV,
$E_7   = 544$ meV,
$E_8   = 549$ meV,
$E_9   = 577$ meV,
$E_{10}= 534$ meV,
$E_{11}= 589$ meV,
$E_{12}= 576$ meV.
For all other barriers $E_\mathrm{far}=588$ meV.
}
\label{fig:barriers}
\end{figure}

The only allowed motion is the exchange of the vacancy with one of the
neighboring atoms. The exchange rate depends on the local environment,
i.e. on the relative position of the vacancy and the impurity atom.
This takes into account the effect of the lattice stress induced by the
tracer atom on the energy landscape observed by the vacancy. Each rate
was assumed to be simply proporional to the Boltzmann factor
$e^{-\frac{\Delta E}{k_BT}}$, where $\Delta$E is the activation energy
for the considered diffusion step and $k_BT$ is the thermal energy at
temperature T. The
diffusion barriers of the vacancy were calculated with the EAM method
\cite{eam84}. We used the interatomic potentials proposed by Finnis and
Sinclair \cite{finnis84}, where the metallic cohesion is taken into
account via the second moment approximation to the tight-binding model.
The calculation was done on a $15\times 15\times 15$ slab of copper. The
system with different relative positions of an In atom and a surface
vacancy was fully relaxed until a threshold force of
$1.875\cdot10^{-12}$ N on the atoms. The barriers were taken at the
point halfway between the stable sites.  See figure~\ref{fig:barriers}
for the values for the barriers.  Since the barrier differences are
large compared to $k_\mathrm{B} T$, the difference in jump probabilities
are extremely large (see below for typical values).

When the vacancy reaches the perimeter of the lattice, its random walk is
terminated, corresponding to the physical process of its recombination
at surface steps. During its lifetime, the vacancy displaces atoms along
its path, many of them multiple times. Thus also the tracer atom can
end up displaced from its original position at the time of recombination of the
vacancy. Averaged over the random walks of many independent vacancies, this yields a
probability distribution of the different displacement vectors that the tracer
atom can make as a result of its encounter with a single vacancy. Due to the
boundary conditions, introduced by the finite size of the lattice and due to
the distribution of exchange rates, an analytic solution to this problem is no
longer possible.

When treating the above model numerically, we separate the motion of the
vacancy and the tracer atom, as has been performed also in some
of the analytical treatments referred to in section \ref{sec:tracer}.
In our case of a finite lattice, this separation introduces an approximation,
which is valid only if the tracer atom is relatively close to the middle of
the lattice. First, we calculate the probabilities that the vacancy,
released at one atomic spacing from the tracer, returns the first time
to the tracer from equal (${p_\mathrm{eq}}$), perpendicular
($p_\mathrm{perp}$) or opposite ($p_\mathrm{opp}$) directions; we also calculate
the probability of its recombination ($p_\mathrm{rec}$) at the perimeter
instead of returning to the tracer. Knowing these return and recombination
probabilities, we turn to the motion of the tracer
atom, which performs a biased random walk of finite length. The direction of
each step with respect to the previous one, and the probability that
this was the last step, are obtained from the return and the recombination
probabilities. With this method we lose all temporal information about the
random walk of the tracer, but 
the individual steps of the tracer atom are orders of magnitude too
fast to be resolved with the STM, and therefore the STM observations are
insensitive to this information anyway.

In practice, both the return probabilities and the motion of the tracer
atom are obtained with direct evaluation of probabilities, which has
better convergence properties than Monte-Carlo-type methods.

As an illustration of this enumeration method, let us consider the
computation of the vacancy
return probabilities. We assign a variable to each lattice site, which
measures the probability that the vacancy after $s$ atomic steps is at
that site, while it has not exchanged with the tracer yet. Initially all
probabilities are zero except at (1,0) where it is unity: we release the
vacancy from here. The boundary acts as a trap for the vacancy, as well
as the site (representing the tracer) at the origin. For each atomic
step of the vacancy we update the site variables parallel by
distributing their probability to the four neighbors according to the
respective exchange probabilities. As the probability flows into the
trap at the origin, we record the cumulative flow in each of the four
directions leading to that site, which gives the return probabilities at
the end. This iteration converges fast, and the convergence can be
measured by the sum of the probabilities still on the lattice. The other
computation, the motion of the tracer atom, is similar but slightly more
complex. In that computation we assign a variable to each incoming edge
of each site, which measure the probability that after $s$ steps ---
each corresponding to a vacancy return --- the tracer is at the given
site and that it arrived from the given direction. In addition, each site has
a variable which accumulates the probability that the tracer become
immobile at that site. These probabilities for the tracer arrival and 
immobilization are updated iteratively according to the previously
obtained vacancy return probabilities.

We first tested our model on the case of unbiased vacancy diffusion,
which would correspond to infinite temperature. The shape of the tracer
distribution was similar to the experimentally observed one, but to
achieve quantitative agreement the only remaining parameter of the model
--- the lattice size $l$ --- had to be tuned to astronomical sizes. This
clearly shows that taking equal barriers is an unrealistic
oversimplification for the In/Cu(001) system.

Using the EAM barriers, for typical parameters $T=320$ K and $l=401$ the
values for the return probabilities were 
$p_\mathrm{eq}  =1-2.4\times10^{-7}$, 
$p_\mathrm{perp}=1.1\times10^{-7}$, 
$p_\mathrm{opp} =4.2\times10^{-9}$, and the recombination probability
$p_\mathrm{rec} =1.1\times10^{-8}$.
These values depend weakly on $l$ [e.g. the dependence of the mean
square displacement, calculated later in this paper from the return
probabilities, is logarithmic: $\langle r^2\rangle\propto\log (l/l_0)$].
This is a consequence of the fact that two is the marginal dimension for
the return problem of the random walker. In higher-dimensional space
the vacancy does not necessarily return, the return probabilities are
asymptotically independent of the lattice size, and the final
distribution of the tracer --- also independent of lattice size for
large lattices --- takes the following form, e.g. in three dimensions
\cite{newman99}: $p(\mathbf{r}) = C_1 \delta( \mathbf{r}) + C_2
r^{-1}\exp(-\Lambda r)$.

\begin{figure}	%%% Fig lattice
\centering\epsfig{file=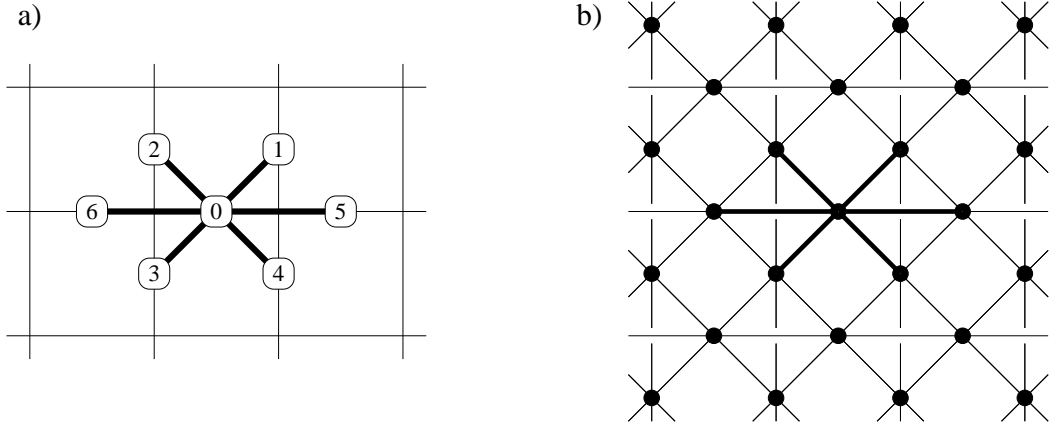,width=\textwidth}
\caption{a) The bond-to-bond steps of the vacancy-tracer pair. If we
view the vacancy-tracer pair to live on the center of each bond,
indicated by the labels, then the pair hops between these sites. The
atomic lattice is drawn with thin lines.  b)~The lattice (made of bonds
of the atomic lattice) on which the vacancy-tracer pair walks. The
neighbors that can be reached by a single step from the site at the
center are shown with thick lines. This lattice can be considered as a
square lattice, rotated by $45^\circ$ with respect to the atomic
lattice, with some extra diagonal bonds. Note that this is not a Bravais
lattice.
}
\label{fig:lattice}
\end{figure}

For the case of an indium impurity in a Cu(001) lattice, the diffusion barrier
for a vacancy exchange with the indium atom is considerably lower than all
other barriers. Therefore, in most cases the vacancy returns from the
direction of its previous departure, and the individual moves of
the tracer atom are strongly anti-correlated. Both the
numerical and the theoretical treatment are simplified significantly if we
do not have to follow the large number of ineffective ``back and forth''
exchanges of the vacancy and the tracer atom. For this purpose, consider
the small probability $\epsilon$ that the
vacancy does {\em not\/} return from the same, equal direction. Thus
${p_\mathrm{eq}} = 1-\epsilon$. In the case of indium in Cu(001) the EAM
calculations yield $\epsilon = 2.4\cdot10^{-7}$ (see above). We now define
\begin{eqnarray}
\hat p_\mathrm{perp} &=& p_\mathrm{perp} / \epsilon \nonumber \\
\hat p_\mathrm{opp}  &=& p_\mathrm{opp}  / \epsilon \label{eq:hat} \\
\hat p_\mathrm{rec}  &=& p_\mathrm{rec}  / \epsilon \nonumber
\end{eqnarray}
and have 
\begin{equation}
2\hat p_\mathrm{perp} + \hat p_\mathrm{opp} + \hat p_\mathrm{rec} = 1.
\label{eq:sum}
\end{equation}
If we represent the quasi-bound state of the rapidly exchanging (on
average $1/\epsilon$ times) vacancy and impurity atom with the position of a {\em bond\/}
of the original lattice, then the vacancy-tracer pair walks on the bonds
of the original lattice. The pair steps to each of the four perpendicular
bonds with probability $\hat p_\mathrm{perp}/2$, and to each of the two
parallel ones with $\hat p_\mathrm{opp}/2$, see figure~\ref{fig:lattice}.
(The factors $1/2$ reflect the probability for the vacancy to escape either
at the right or the left side of the quasi-bound position, which is $\frac{1}{2}$
for both sides in the limit of vanishing $\epsilon$. The advantage
of this approach is twofold: the path of the tracer is made of fewer
steps (beneficial for numerics), and the bond-to-bond steps are now
independent (beneficial for theoretical treatment).

\begin{figure}	%%% Fig jvd
\centering\epsfig{file=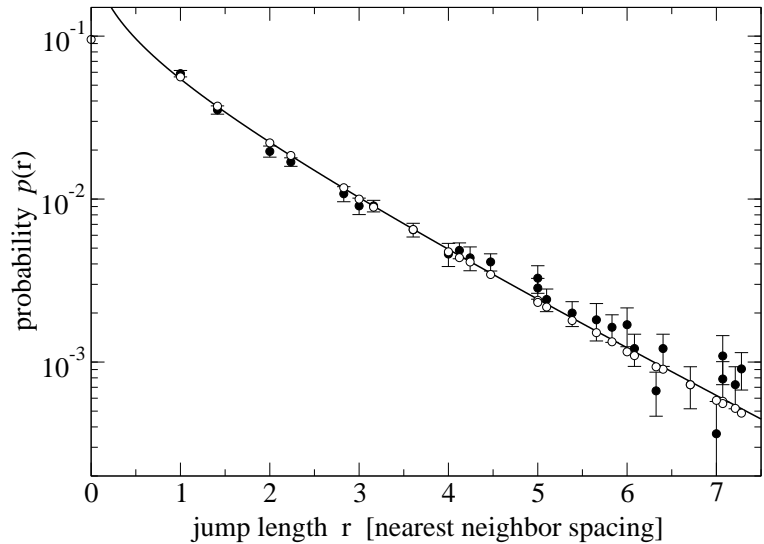,width=10cm}
\caption{The probabilities of the jump lengths of the tracer atom for
$T=320$ K and $l=401$ lattice spacings. Filled circles correspond to
experimental values (measured at this temperature and terrace size),
open circles are from the model described in the text, and the solid
curve is the continuum solution described in
section~\ref{sec:continuum}. The data shows no significant directional
structure: the dependence on the {\em length\/} of the jumps is
monotonic with good approximation. (Each dataset is normalized
separately such that the probabilities corresponding to a subset of the
jump vectors, $1 \le |\mathbf{r}| \le 6$, add up to unity. These are the
probabilities that are determined with good accuracy in the experiment.)
}
\label{fig:jvd}
\end{figure}

Using this, the tracer atom described as if it forms a pair with the
vacancy on one of the bonds adjacent to its original site, walks on the
bonds lattice, and at the end (which happens with probability $\hat
p_\mathrm{rec}$ after each step) it is released at either end of the
last visited bond. Results for the probabilities of the different jump
lengths (beginning-to-end vectors of these paths) are shown on
figure~\ref{fig:jvd}. Note, that the model calculations in
figure~\ref{fig:jvd} contained no adjustable parameters.

\begin{figure}	%%% Fig probzero
\centering\epsfig{file=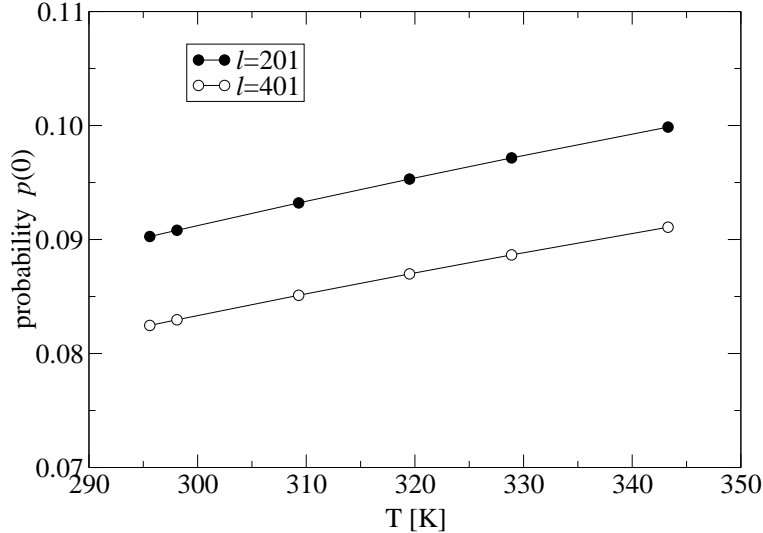,width=10cm}
\caption{The temperature dependence of the probability that a tracer
atom had an encounter with a vacancy but its net displacement was zero,
plotted for two different lattice sizes.  The dependence is weak.
}
\label{fig:probzero}
\end{figure}

A general advantage of numerical modeling is that we have access to
quantities which are difficult or impossible to measure experimentally.
One example in our simulation is the probability that a tracer atom had
an encounter with a vacancy, but its net displacement was zero.  The
temperature dependence of this quantity is plotted on
figure~\ref{fig:probzero}.  Since the dependence is weak, the assumption
in the experimental measurements to associate In-vacancy encounters with
visible (non-zero) jumps of the In atom is justified.

For a given set of diffusion barriers, our model has two parameters: the
temperature and the lattice size. When we compare results with
experiments, both can be independently obtained, and in principle there
are no adjustable parameters. For example the case of $T=320$ K
(figure~\ref{fig:jvd}) --- using the distance to the nearest step in the
experiment as lattice size in the numerical model --- gave a good match
with the experiment, but for the measurements at other temperatures we
had to adjust the lattice size to obtain good agreement.  Although the
best fit lattice size in some cases was a factor of 2-3 smaller than the
measured distance to the nearest step, the change in the return
probabilities was much smaller, as they depend logarithmically on the
lattice size. Undetected defects --- acting as trap for the vacancies
--- closer to the In atom than the nearest step could also be accounted
for this difference.

%%%%%%%%%%%%%%%%%%%%%%%%%%%%%%%%%%%%%%%%%%%%%%%%%%%%%%%%%%%%%%%%%%%%%%%%

\section{Continuum model}
\label{sec:continuum}

% * the need for continuum model: characterize with few parameters
%   drawbacks: small lattice etc
% * from which point we do this
% * derivations
% * set parameters
% * results

In the previous section we solved numerically the version of the tracer
diffusion problem which was relevant for the In/Cu(001) experiments.
Although this already enables full comparison, the numerical solution
has the disadvantage that it cannot be described with a few parameters.
In this section we develop a simple continuum description, where the
overall shape of the jump length distribution is described with a single
parameter.

We use our previous results for the return and recombination
probabilities of the vacancy, and consider the random walk of the
tracer-vacancy pair on the bond lattice.
Let $\varrho(\mathbf{r},n)$ denote the probability that the
tracer-vacancy pair is at position $\mathbf{r}$ and at instance $n$,
where $n$ counts the number of steps the tracer-vacancy pair takes.
Since the subsequent steps of the pair are independent, we can write an
effective diffusion equation for the evolution of
$\varrho(\mathbf{r},n)$:
\begin{equation}
\frac{\partial\varrho(\mathbf{r},n)}{\partial n} = D_\mathrm{eff}
\nabla^2 \varrho - c\varrho \,.
\label{eq:diff}
\end{equation}
The first term on the right hand side corresponds to the steps the pair
takes on the bond lattice, here $D_\mathrm{eff}$ denotes the mean square
displacement per step of the pair.  The second term corresponds to the
recombination of the vacancy\footnote{Note that since $n$ counts the
number of steps of the vacancy-tracer pair on the bond lattice, the term
$-c\varrho$ does not imply that the vacancy can recombine at any lattice
site --- in fact it recombines at terrace steps, between subsequent
returns to the In atom.}; at this point the pair breaks up. In the
continuum approximation for space and $n$, the solution for a
Dirac-delta initial condition at the origin is 
\begin{equation}
\varrho(\mathbf{r},n) = \frac{1}{4\pi D_\mathrm{eff}n}
\exp\left(-\frac{r^2}{4D_\mathrm{eff}n} - c n\right) \,.
\end{equation}
The final distribution of the tracer atom after the vacancy recombined
is obtained by the integration of the loss term in Eq.~(\ref{eq:diff}):
\begin{equation}
p(\mathbf{r}) = \int_0^\infty c \varrho(\mathbf{r},n) \d n
= \frac{1}{2\pi} \frac{c}{D_\mathrm{eff}} \, \mathrm{K}_0
  \left(\frac{r}{\sqrt{D_\mathrm{eff}/c}}\right) \,,
\end{equation}
where $\mathrm{K}_0$ is the modified Bessel function of order 0.  The
functional form of this solution is similar to that of the infinite
lattice system in section~\ref{sec:tracer}, in spite of the differences
introduced by the finite vacancy lifetime and the different boundary
conditions.

The mean square displacement $\langle r^2\rangle$ is directly
proportional to the square of the width of the Bessel solution, apart
from lattice corrections. We can determine the width of the Bessel
solution from the parameters $D_\mathrm{eff}$ and $c$, which are
obtained from the return probabilities in the Appendix:
\begin{equation}
\langle r^2\rangle \, \propto \, \frac{D_\mathrm{eff}}{c} =
\frac{\hat p_\mathrm{perp}+ \hat p_\mathrm{opp}} {4 \hat p_\mathrm{rec}}
\,.
\end{equation}

This continuum solution is shown in figure~\ref{fig:jvd}. It closely
follows the numerical solution of the model, even for relatively small
distances from the origin, where one would expect stronger lattice
effects.

%%%%%%%%%%%%%%%%%%%%%%%%%%%%%%%%%%%%%%%%%%%%%%%%%%%%%%%%%%%%%%%%%%%%%%%%

\section{Summary}
\label{sec:summary}

In this paper we described a model for vacancy mediated tracer diffusion
on a finite lattice with absorbing boundaries for the vacancy. These
boundary conditions were appropriate to model the vacancy mediated
diffusion of In atoms embedded in the top layer of Cu(001) terraces. In
addition to the numerical solution of the discrete model, we set up a
simple continuum formulation of the model. The spatial distribution of
the tracer atom in the continuum solution has a modified Bessel function
profile. This form of non-Gaussian distribution is typical for tracer
diffusion assisted by other diffusing particles. In order to enable a
quantitative comparison with the STM measurement of the In/Cu(001)
system, we introduced modified vacancy diffusion rates near the
In atom, calculated with the EAM method. The modified rates affect the
width of the Bessel function, without changing the functional form of
this characteristic distribution.

\begin{ack}
We thank L. Niesen and M. Rosu from the University of Groningen for
providing us with the computer code for the EAM calculations.  This work
is part of the research program of the ``Stichting voor Fundamenteel
Onderzoek der Materie (FOM),'' which is financially supported by the
``Nederlandse Organisatie voor Wetenschappelijk Onderzoek (NWO).''
\end{ack}

%%%%%%%%%%%%%%%%%%%%%%%%%%%%%%%%%%%%%%%%%%%%%%%%%%%%%%%%%%%%%%%%%%%%%%%%

%%%%%%%%%%%%%%%%%%%%%%%%%%%%%%%%%%%%%%%%%%%%%%%%%%%%%%%%%%%%%%%%%%%%%%%%

\begin{appendix}
\section{Obtaining $D_\mathrm{eff}$ and $c$ from the return
probabilities}
\label{sec:appendix}

In this Appendix we calculate the parameters of the continuum model,
$D_\mathrm{eff}$ and $c$, from the return and recombination
probabilities in Eq.~(\ref{eq:hat}).  The probability $P_0^{(n+1)}$ that
the vacancy-tracer pair is at site ``0'' (not necessarily the origin)
after $n+1$ steps of the pair is contributed from sites
$1,2,\,\ldots\,,\,6$ (see figure~\ref{fig:lattice}a; no upper index
refers to probabilities after $n$ steps):
\begin{equation}
P_0^{(n+1)} = \frac{\hat p_\mathrm{perp}}{2}(P_1+P_2+P_3+P_4)
  + \frac{\hat p_\mathrm{opp}}{2}(P_5+P_6) \,.
\label{eq:P0}
\end{equation}

During one step in $n$, the change in $P_0$ is (using Eq.~\ref{eq:sum})
\begin{eqnarray}
P_0^{(n+1)} - P_0 =&& \frac{\hat p_\mathrm{perp}}{2}
  (P_1+P_2+P_3+P_4-4P_0) \nonumber \\
 && + \frac{\hat p_\mathrm{opp}}{2}(P_5+P_6-2P_0) - \hat p_\mathrm{rec}
    P_0 \,.
\end{eqnarray}
The finite difference on the left hand side approximates the derivative,
and the terms on the right hand side can be collected to form a lattice
approximation to the Laplacian:
\begin{equation}
\nabla^2P=a_\mathrm{perp}^{-2}(P_1+P_2+P_3+P_4-4P_0)
\label{eq:lap1}
\end{equation}
and 
\begin{equation}
\partial^2_x P = a_\mathrm{opp}^{-2}(P_5+P_6-2P_0) \approx \frac{1}{2}
\nabla^2P \,,
\label{eq:lap2}
\end{equation}
where the distances $a_\mathrm{perp}=1/\sqrt{2}$ and $a_\mathrm{opp}=1$
are in units of nearest neighbor spacing of the atomic lattice. In the
approximation of Eq.~(\ref{eq:lap2}) we assumed on average equal second
derivatives in all directions. After these substitutions
\begin{equation}
\frac{\partial P_0}{\partial n}
= \underbrace{\frac{\hat p_\mathrm{perp}+\hat
 p_\mathrm{opp}}{4}}_{D_\mathrm{eff}} \nabla^2 P - \underbrace{\hat
 p_\mathrm{rec}}_{c} P_0 \,,
\end{equation}
and the coefficients $D_\mathrm{eff}$ and $c$ can be read off.

\end{appendix}

\end{document}